\documentclass[aps,prb,preprint,superscriptaddress,showpacs]{revtex4-1}
\usepackage[pdftex]{graphicx}
\usepackage{color}
\usepackage{amsmath}

\usepackage[final]{changes}

\newcommand{\n}{$_{\mathrm n}$}
\newcommand{\nc}{n$_{\mathrm C}$}

\usepackage{xr}
\newcommand{\ICMM}[0]{{
Instituto de Ciencia de Materiales de Madrid, CSIC,
Cantoblanco, 28049 Madrid, Spain}}

\newcommand{\QUIM}[0]{{
Departamento de Pol\'{\i}meros y Materiales Avanzados: F\'{\i}sica, Qu\'{\i}mica y Tecnolog\'{\i}a, Facultad de Qu\'{\i}mica,
Universidad del Pa\'{\i}s Vasco UPV/EHU,
Apartado 1072, 20080 Donostia-San Sebasti\'an, Spain}}

\newcommand{\DIPC}[0]{{
Donostia International Physics Center,
Paseo Manuel de Lardiz\'abal 4, 20018 Donostia-San Sebasti\'an, Spain}}

\newcommand{\CFM}[0]{{
Centro de F\'{\i}sica de Materiales CFM/MPC (CSIC-UPV/EHU),
Paseo Manuel de Lardiz\'abal 5, 20018 Donostia-San Sebasti\'an, Spain}}

\newcommand{\IKER}[0]{{
IKERBASQUE, Basque Foundation for Science, 48013 Bilbao, Spain}}

\newcommand{\IMDEA}[0]{{
IMDEA Nanociencia, Campus de Cantoblanco, 28049 Madrid, Spain}}

\newcommand{\UAM}[0]{{
Departamento de F\'{\i}sica de la Materia Condensada,
Instituto Nicol\'as Cabrera, and Condensed Matter Physics Center (IFIMAC),
Universidad Aut\'onoma de Madrid, 28049 Madrid, Spain}}

\newcommand{\ALBA}[0]{{
ALBA Synchrotron Light Source, Cerdanyola del Valles, 08290 Barcelona, Spain}}

\newcommand{\COMPLU}[0]{{
Departamento de F\'{\i}sica de Materiales and
Instituto Pluridisciplinar, Universidad Complutense de Madrid, Ciudad Universitaria 28040, Madrid, Spain}}

\begin{document}

\author{M. Blanco-Rey}
\affiliation{\QUIM}
\affiliation{\DIPC}
\email{maria.blanco@ehu.es}
\author{P. Perna}
\affiliation{\IMDEA}
\email{paolo.perna@imdea.org}
\author{A. Gudin}
\affiliation{\IMDEA}
\author{J.M. Diez}
\affiliation{\IMDEA}
\affiliation{\UAM}
\author{A. Anad\'on}
\affiliation{\IMDEA}
\author{Leticia de Melo Costa}
\affiliation{\IMDEA}
\affiliation{\ALBA}
\author{Manuel Valvidares}
\affiliation{\ALBA}
\author{Pierluigi Gargiani}
\affiliation{\ALBA}
\author{Alejandra Guedeja-Marron}
\affiliation{\IMDEA}
\affiliation{\COMPLU}
\author{Mariona Cabero}
\affiliation{\COMPLU}
\author{M. Varela}
\affiliation{\COMPLU}
\author{C. Garc\'{\i}a-Fern\'andez}
\affiliation{\CFM}
\affiliation{\DIPC}
\author{M.M. Otrokov}
\affiliation{\IKER}
\affiliation{\CFM}
\affiliation{\DIPC}
\author{J. Camarero}
\affiliation{\IMDEA}
\affiliation{\UAM}
\author{R. Miranda}
\affiliation{\IMDEA}
\affiliation{\UAM}
\author{A. Arnau}
\affiliation{\QUIM}
\affiliation{\CFM}
\affiliation{\DIPC}
\author{J.I. Cerd\'a}
\affiliation{\ICMM}

\title{Origin of the Large Perpendicular Magnetic Anisotropy in
Nanometer-thick Epitaxial Graphene/Co/Heavy Metal Heterostructures}

\date{\today}

\begin{abstract}
A combination of theoretical modelling and experiments reveals the origin of the large
perpendicular magnetic anisotropy (PMA) that appears in nanometer-thick epitaxial Co films
intercalated between graphene (Gr) and a heavy metal (HM) substrate, as a function of the Co thickness.
High quality epitaxial Gr/Co\n/HM(111) (HM=Pt,Ir) heterostructures are grown by intercalation below
graphene, which acts as a surfactant that kinetically stabilizes the pseudomorphic
growth of highly perfect Co face-centered tetragonal ($fct$) films,
with a reduced number of stacking faults as the only structural defect
observable by high resolution scanning transmission electron microscopy (HR-STEM).
Magneto-optic Kerr effect (MOKE) measurements show that such heterostructures present PMA up to
large Co critical thicknesses of about 4~nm (20~ML) and 2~nm (10~ML) for Pt and Ir substrates, respectively,
while X-ray magnetic circular dichroism (XMCD) measurements show an
inverse power law of the anistropy of the orbital moment with Co thickness,
reflecting its interfacial nature, 
that changes sign at about the same critical values.
First principles calculations show that, regardless of the presence of graphene,
ideal Co $fct$ films on HM buffers do not sustain PMAs beyond around 6~MLs 
due to the in-plane contribution of the inner bulk-like Co layers. 
The large experimental critical thicknesses sustaining PMA can only be retrieved
by the inclusion of structural defects that promote a local $hcp$ stacking
such as twin boundaries or stacking faults.
Remarkably, a layer resolved analysis of the orbital momentum anisotropy
reproduces its interfacial nature, and reveals
that the Gr/Co interface contribution is comparable to that of the Co/Pt(Ir).
\end{abstract}

\maketitle

\section{Introduction}
\label{sec:introduction}

Heterostructures with large perpendicular magnetic anisotropy 
(PMA)~\cite{Dieny.rmp2017,bib:Carcia88,bib:Su09,bib:winkler15,Yang.nl2016,Rougemaille.apl2012,Zhao.acsn2018,Xie.advmat2019,Zhang.acsn2019,Weinan.prl2020}  
are a key ingredient in the emerging field of spin orbitronics~\cite{Soumyanarayanan.nat2016}, 
aimed at the development of functional, high-speed, low-energy consumption nanodevices~\cite{Manchon.rmp2019}. 
Maximizing the PMA is essential to downscale the size of data storage spintronic devices such as 
spin transfer torque magnetic random access memories (STT-MRAM)~\cite{Manchon.rmp2019}. 
In metallic multilayer heterostructures, the effective PMA is determined by the morphology 
plus an intricate interplay of structural and electronic effects, the latter being essentially 
dependent on the crystal field and the spin orbit interaction (SOI) strength~\cite{Dieny.rmp2017}. 
This is often enhanced by growing alternate ultrathin layers of magnetic (e.g., Co or Fe) 
films and heavy metals (HM), like Pt or Ir, which induce strong SOI at the interfaces by a 
proximity effect~\cite{bib:Carcia88}. However, as the thickness of the constituent layers increases, 
the PMA is severely reduced, since the the morphology and the structure of the magnetic films may degrade, 
the weight of the interfaces is reduced and the low-dimensional behavior is lost.

Design, fabrication, and characterization of multilayer structures with large PMA is a 
thriving open research area. The growth conditions determine the resulting 
morphology, i.e., two-dimensional \emph{vs.} three-dimensional films, structural 
details~\cite{bib:winkler15}, such as the crystalline perfection and the stacking 
sequence of atomic planes in each constituent layer, which may include dislocations, 
stacking faults (SFs)~\cite{Ascolani.ss1996,bib:vazquez00,bib:doi15} or twin boundaries (TBs), 
as well as strain or compositional disorder due to intermixing. Among all of them, 
a three-dimensional morphology rapidly cancels interfacial effects in thin ferromagnetic films. 
Recently, a new technique~\cite{bib:ajejas18, bib:ajejas20} has been developed to grow high quality 
Co(111)/Pt(111) heterostructures by intercalation of Co atoms in Gr/Pt(111) systems, 
with graphene playing the role of a surfactant, a similar behaviour as in intercalation on
a Ir(111) buffer \cite{bib:rougemaille12,bib:decker13,bib:gargiani17,bib:vlaic18}. 
The result is an atomically flat film of 
highly perfect, pseudomorphic Co in a face centered tetragonal ($fct$) crystalline structure, 
laterally expanded to adjust to the lattice parameter of Pt, which keeps a giant PMA for unusually 
large thickness of the nanometer-thick Co film~\cite{bib:ajejas18}.

In this work, we unravel the origin of this large PMAs in epitaxial $fct$ Co 
films intercalated between graphene and HM substrates, 
i.e. Gr/Co\n/HM(111) (HM = Pt, Ir and n being the number of Co layers).
Magneto-optic Kerr effect (MOKE) measurements reveal that these heterostructures develop PMA 
up to unusually large Co thicknesses of about 4~nm ($\simeq$20~MLs) and 2~nm ($\simeq$10~MLs) 
for Pt and Ir substrates, respectively. Consistently, X-ray magnetic circular dichroism (XMCD) 
experiments evidence that at the same critical thicknesses the orbital momentum anisotropy is 
switched from out-of-plane to in-plane. Detailed state-of-the-art \emph{ab initio} calculations 
within density functional theory (DFT), converged with unprecedented 
accuracy for a number of structural scenarios (Co film thicknesses in 
the $1-20$~MLs range, different Gr moir\'e patterns, inclusion
of TBs/SFs in the Co film, and intermixing at the interfaces), 
allow breaking down the contributions responsible for this behavior, 
the most important of them coming from local $hcp$ stackings in the Co film induced by TBs or SFs. 
Our high-resolution scanning transmission electron microscopy (HR-STEM) measurements indeed confirm 
the presence of SFs in both Pt- and Ir-based systems that inevitably appear associated to atomic 
steps at the Co/HM interface. Extensive theoretical modelling of various TBs in the 
$fcc$ stacked intercalated Co films shows that these defects play a fundamental role in sustaining 
the PMA up to the observed large Co thicknesses. Finally, we find a sizable orbital momentum 
anisotropy induced by graphene in the two outer Co atomic planes, which, in spite of the much 
weaker SOI in graphene, appears to be comparable to that at the Co/Pt(Ir) interface.

\section{Results and Discussion}
\label{sec:results}

\begin{figure}
\includegraphics[width=.7\columnwidth]{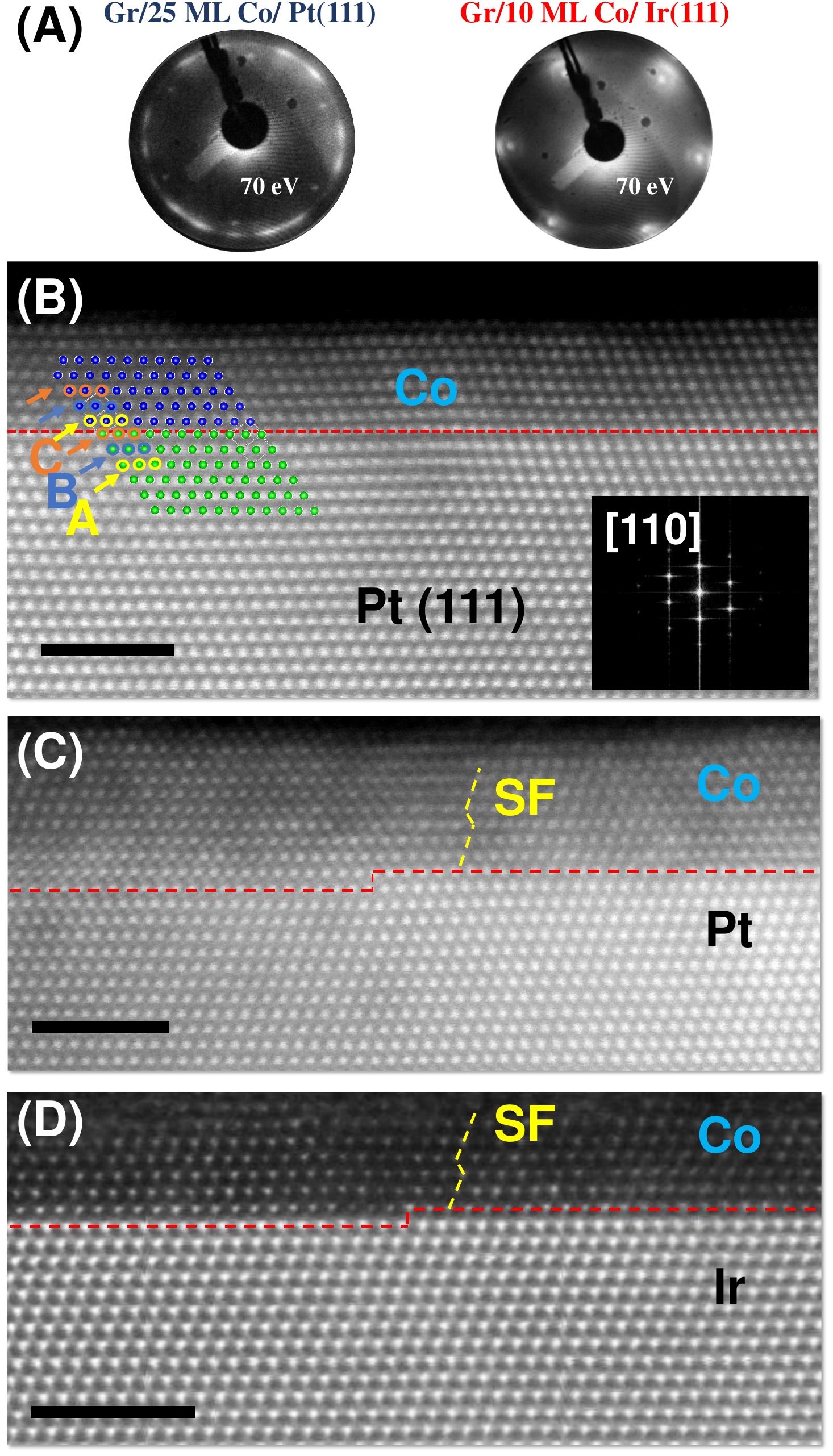}
\caption{\label{fig:sample_characterization} 
(Caption on next page)
}
\end{figure}

The structural characterization of the epitaxial Gr/Co/HM(111) (HM=Pt, Ir) heterostructures 
is shown in Fig.~\ref{fig:sample_characterization}. The intercalated Co films are atomically 
flat as revealed by STM~\cite{bib:ajejas20}.
The LEED patterns observed after intercalation (Fig.~\ref{fig:sample_characterization}(A)) 
are identical to the ones seen for pristine Gr/Pt(111) and Gr/Ir(111) corresponding to the 
moir\'e patterns also visualized for both substrates with STM~\cite{bib:sutter09,bib:ndiaye08,bib:decker13}
and different from the $(1\times 1)$ pattern observed for Gr/Co(0001). This confirms that 
(i) graphene is floating on top of the Co films, i.e. Co is intercalated 
between the HM(111) buffer layer and graphene; and 
(ii) the Co films are pseudomorphic with the Pt and Ir substrates, i.e. laterally expanded 
to adjust to the lattice parameter of the substrate. 
The presence of Co underneath graphene is verified by XPS, as demonstrated elsewhere~\cite{bib:ajejas20}. 
The high angle annular dark field (ADF) STEM images, 
see also Supplementary Material (SM) Figure~\ref{SM-fig:microstr}, 
reveal the predominant $fcc$ stacking of the Co layers intercalated and the pseudomorphic 
arrangement of the Co film, as well as the presence of stacking faults (SFs) at the Co layer 
near the Pt (Ir) steps underneath. The lateral (tensile) strain induced by the HM substrate 
results in a tetragonal distortion~\cite{bib:cerda93} of the $fcc$ Co 
towards an $fct$ structure.

\addtocounter{figure}{-1}
\begin{figure}[!t]
\caption{
{\bf Structural and microscopic characterization of epitaxial Gr/Co$_{\mathrm n}$/HM heterostructures.}
(A) Representative LEED patterns acquired at 70\,eV on the indicated samples after Co deposition
and intercalation. The moir\'e superstructures of Gr on Pt(111) and Ir(111) are preserved after
Co intercalation.
%
%
(B), (C) and (D) High angle annular dark field (HAADF) high resolution STEM images of the interfaces
upon intercalation of 10ML of Co on Gr/Pt(111) and Gr/Ir(111). Panel (B) corresponds to Co on Gr/Pt(111.
The sketch overlaid across the HM/Co interface highlights the ABCABC atomic plane stacking sequence of the
$fct$ lattice, marked with colored arrows, and the pseudomorphic arrangement of the Co film.
Panels (C) and (D) show high magnification images of stacking faults in 10\,MLs thick Co over 
Pt(111) (C) and Ir(111) (D) buffers grown epitaxially onto SrTiO$_3$ (STO) (111) substrates.
Stacking faults can be observed in both samples, marked with yellow dashed lines.
The approximate interface positions are marked with red dashed lines.
Occasional atomic steps are visible. The scale bars represent 2\,nm.
The inset shows the fast Fourier transform of the STEM image, acquired along the [110] projection,
revealing the high crystalline coherence between Co and HM layers.
}
\end{figure}

\begin{figure}
\includegraphics[width=0.8\columnwidth]{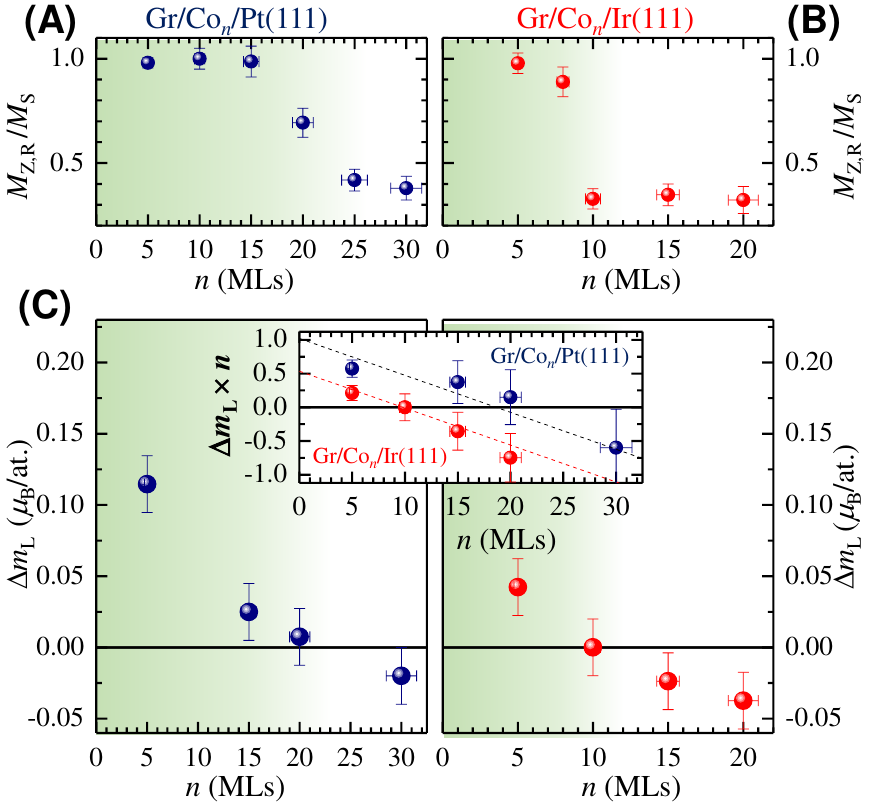}
\caption{\label{fig:mokexmcd}
{\bf Quantified magnetic parameters of epitaxial Gr/Co/HM heterostructures.}
Room-temperature thickness-dependent remanence (top graphs), and anisotropy of orbital moment
(bottom) of Gr/Co\n/Pt(111) (A) and Gr/Co\n/Ir(111) (B) samples.
$M_{Z,R}/M_S$ is the remanent perpendicular magnetization normalized to the saturation magnetization.
Symbols are the data derived from polar-MOKE and XMCD measurements, such as
the ones shown in Supplementary Material Figures~\ref{SM-fig:moke} and \ref{SM-fig:xmcd1}, respectively.
(C) The anisotropy of the orbital moment ($\Delta m_L$) is computed from the difference of orbital
moments derived from sum-rule analysis at normal (NI) and grazing (GI) incidence geometries
($\Delta m_L = m_{L,\mathrm{NI}} - m_{L,\mathrm{GI}}$), assuming a number of
Co-$d$ holes of 2.49.
The shadowed areas in the graphs highlight the corresponding PMA critical Co
thickness, that is \nc=20~ML and \nc=10~ML for Pt and Ir buffers, respectively.
Notice the coincidence of \nc\ and the change of sign of $\Delta m_L$. The inset
displays the corresponding anisotropy of orbital
moment times Co thickness versus Co thickness. Notice the similar linear slope
(bulk contribution) and different vertical axis intercept (interface contribution), 
twice as large in the case of Gr/Co/Pt.}
\end{figure}

A well-defined perpendicular magnetic anisotropy (PMA), i.e.,
out-of-plane magnetization easy axis, in Gr/Co$_{\mathrm{n}}$/HM
heterostructures has been identified at 300\,K below a critical Co
thickness ($n_{\rm C}$) by polar magneto-optic Kerr effect
(polar-MOKE) and X-ray Magnetic Circular Dichroism (XMCD)
measurements. From the polar-MOKE hysteresis loops with the
external field applied to the surface normal, the remanence and
saturation magnetizations, $M_{Z,R}$ and $M_S$, respectively,
are determined as a function of Co thickness ($n$) for samples grown on Pt and Ir 
(see Supplementary Material (SM) Fig.~\ref{SM-fig:moke}). 
The corresponding ratios $M_{Z,R}/M_S$, are shown in Fig.~\ref{fig:mokexmcd}. 
In the case of the Pt(111) substrate, the hysteresis loops evolve smoothly from a square-shaped loop 
with large coercive field, i.e. 100~mT for n$=5$ Co MLs
to an S-shaped loop with reduced remanence values above \nc. 
Fig.~\ref{fig:mokexmcd}(A) suggests that 
the magnetization switches from out-of-plane to 
in-plane upon growth of between n$=15-25$ Co atomic planes, i.e., $n_{\mathrm C} \sim 20$~ML. 
The hysteresis behavior change is more 
abrupt in the case of Ir(111), where the magnetization reorientation happens at a lower Co thickness 
value around $n_{\rm C} \sim 10$~ML as can be seen in Fig.~\ref{fig:mokexmcd}(B). 

Angular dependence XMCD measurements performed at room temperature (RT) in
Gr/Co$_\mathrm{n}$/HM heterostructures confirm qualitatively the
aforementioned polar-MOKE observations. The comparison between the
dichroism spectra acquired in normal incidence (NI) and grazing
incidence (GI) geometries for the different Co thicknesses
directly shows: i) PMA for $n<n_{\rm C}$ (i.e., larger NI
dichroism signal, as Fig. S3 shows) ii) whereas preferential
in-plane orientation for $n>n_{\rm C}$ (i.e., larger GI
dichroism signal, as depicted the bottom graphs of Fig. S4); iii)
the critical thickness is higher in the case of the Pt buffer,
i.e., for HM=Pt. In addition, iv) the quantitative analysis of the
XMCD spectra provides the microscopic interfacial picture on the
origin of the PMA. 
Sum rules applied to the XMCD spectra~\cite{bib:thole92,bib:stoehr99} recorded at 6\,T 
external magnetic field (see SM Fig.~\ref{SM-fig:xmcd1}) 
provide the projection of the orbital magnetic moments along 
the applied field (same as incident light) direction. 
Following the standard application of the sum rules~\cite{bib:chen95},
Fig.~\ref{fig:mokexmcd}(C) shows the orbital moment difference $\Delta m_L$ between out-of-plane 
and in-plane incidence ($\Delta m_L = m_{L,\mathrm{NI}} - m_{L,\mathrm{GI}}$), 
normalized by the number of holes $n_h=2.49$ in the Co-$3d$ band extracted from the XMCD spectra. 
There is a sign change in this quantity between n$=20$ and n$=30$~MLs of Co for Co/Pt and 
around n$=10$~MLs for Co/Ir. These thicknesses are similar to those that yield the magnetization 
reorientation observed by MOKE. 

According to Bruno~\cite{bib:chappert88,bib:bruno89}, $\Delta m_L$
can be written as the combination of bulk and interfaces contributions, i.e.,
$\Delta m_L = \Delta m_L (\mathrm{Co}_\mathrm{bulk}) + [\Delta m_L(\mathrm{Gr/Co})- \Delta m_L(\mathrm{Co/Pt})]/\mathrm{n}$.
This expression should hold as
long as the interfaces are flat and the Co film is thick enough
for the internal Co layers to be considered as an effective
bulk-like contribution, which includes the defect-free $fct$
contribution and further interfacial contributions, such as
stacking defects, as those observed in
Fig.~\ref{fig:sample_characterization}(B). 
Experimentally, when $\Delta m_L$ is multiplied by n (see inset of 
Fig.~\ref{fig:mokexmcd}(C)), we observe a linear dependence 
with the Co thickness with similar slope for both Pt and Ir buffer cases 
which may be assigned to the effective bulk Co contribution, and different vertical
axis intercepts that correspond instead to the interfacial contributions, twice
as large for Gr/Co/Pt(111) than for Gr/Co/Ir(111). 
A similar effective bulk contribution suggests a similar fraction
of defect-free $fct$ and stacking faults into of the Gr/Co$_\mathrm{n}$/HM
heterostructures grown on a similar STO substrate, for both Pt and
Ir cases, which could be associated with the substrate having on
average the same number of steps. A priori the experimental data
are not sufficient to determine the contribution of each interface
from the vertical axis intercept. Considering a negligible Gr/Co
interfacial contribution for both Pt and Ir cases, the
experimental intercepts can be explained with an interfacial Co/HM
contribution two times larger for the Pt case. In contrast, there
is the observation of the unprecedented very high critical
thickness that cannot be understood without 
the presence of the Gr/Co interface.

\begin{figure}
\includegraphics[width=0.9\columnwidth]{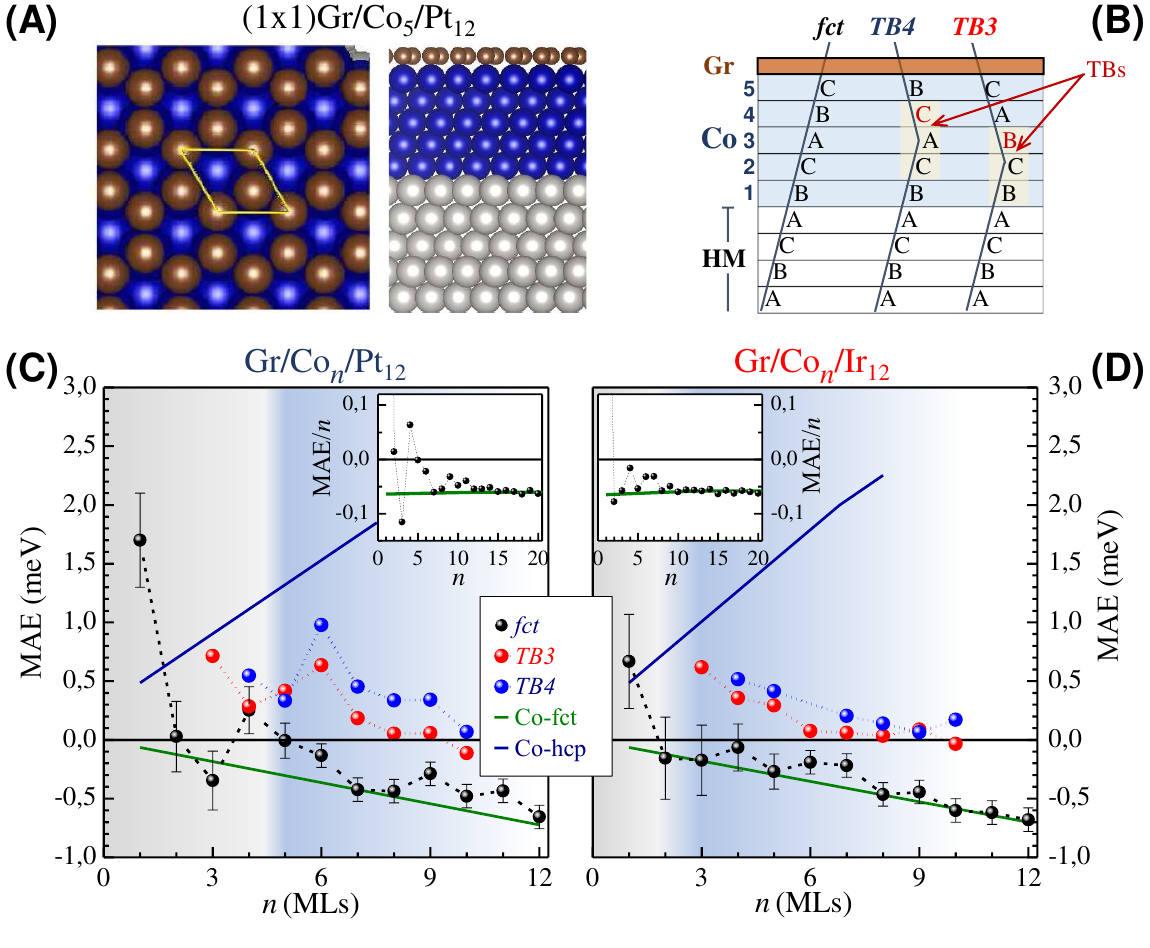}
\caption{\label{fig:mae} 
{\bf Calculated magnetocrystalline anisotropy energy (MAE) for 
($1\times1$)-Gr/Co\n/HM$_{12}$ heterostructures.}
(A) Representative top and side views of the model used for the ideal $fct$
structure with HM=Pt and n=5~MLs.
(B) Scheme of the different layer sequences used, without stacking faults (ideal $fct$) and with one 
twin boundary starting at the third (TB3) or fourth (TB4) layer of the Co film.
(C) and (D) MAE as a function of the Co thickness n calculated for the 
stacking sequences indicated in panel B, in the case of HM= Pt and Ir, 
respectively. Symbols correspond to ($1\times1$)-Gr/Co\n/HM$_{12}$
heterostructures with different stacking sequences (black, red, and blue are 
used for ideal $fct$, TB3 and TB4, respectively), 
whereas continuous lines refer to bulk Co phases ($fct$ and strained $hcp$ 
indicated in dark blue and green, respectively). 
Shadowed areas emphasize the predicted PMA Co critical thickness, \nc, for 
ideal $fct$ films, larger in the case of Pt in comparison with Ir, and its
enhancement when TBs are introduced. 
Insets show the MAEs normalized by the Co thickness, n, for the ideal $fct$ films with n 
up to 20~MLs, in order
to illustrate their convergence towards the bulk Co-$fct$ values indicated by the
horizontal green line.}
\end{figure}

In the following, we rationalize the experimental findings by presenting
DFT-derived magnetic anisotropy energies (MAEs) and orbital magnetic moments
(OMMs) for ideal and defected Gr/Co\n/HM heterostructures 
(see Figures~\ref{fig:mae}(A) and (B)) after including the SOI self-consistently.

{\bf Ideal $fct$ Co films.} Figures~\ref{fig:mae}(C) and (D)
show the calculated MAEs for (1$\times$1)-Gr/Co\n/HM$_{12}$ slabs (black dots 
and lines) as a function of the Co 
thickness, n, assuming an ideal $fcc$ stacking throughout the film
(see SM section~\ref{SM-sec:allmodels} for details on the
relaxed geometries). Inspection of the curves reveals two different
regimes with similar behaviors in both systems (notice that positive values of the MAE correspond to PMA).
At small Co thicknesses, n$< 9$ in Pt and n$< 7$ in Ir
supported films, the anisotropies are highly non-linear presenting multiple oscillations.
Large PMA values are only attained in the ultrathin limit n$=1$, while small
or negligble PMAs also appear at slightly larger thicknesses (n$=2,4-5$ for
Pt and n$=4$ for Ir). In the second region, 
n$\ge9$ (7) for
Pt (Ir), in-plane magnetization has clearly set in and the MAEs
present a quasi-linear behavior with the Co film thickness.
The absence of strong oscillations in this region indicates that the 
individual contributions from the Co/Gr and HM/Co interfaces are essentially 
decoupled, so that the slopes may be associated with the MAE of a tetragonally
distorted Co $fct$ single crystal with the in-plane lattice parameter fixed
to that of the HM (green lines). This is best seen in the insets,
where the MAEs normalized by the number of Co layers, up to n$=20$, are shown
along with that of the corresponding Co bulk $fct$, which is weakly in-plane 
anisotropic. Furthermore,
the coincidence between the two lines, within less than 0.01~meV
for n$\ge12$ (10) in the case of Pt (Ir), 
reveals that, once the film is thick enough and finite quantum size effects are
removed, the net contribution of the two interfaces in both systems is 
negligible. 

Overall, the calculated MAEs are in qualitative agreement with MOKE and
XMCD data, in the sense that they reproduce the switch from out-of-plane
to in-plane as the Co film grows, with the transition ocurring earlier in
Ir than in Pt. However, the calculated critical thicknesses for the MAE switch, \nc,
are notably smaller than the observed values.
Let us recall that in the simulations, apart from ultra-fine $k$-meshes larger
than $70\times70$, we have employed a 12 layer thick HM buffer layer
(see SM section~\ref{SM-sec:bufferthickness}), which is
much larger than those typically used in similar theoretical 
studies~\cite{bib:blonski10,bib:doi15,bib:yang15,bib:steiner16,bib:wolloch17,bib:cabral19},
and, in order to ensure converged MAE values below 0.1~meV,
we have also avoided the use of perturbative approaches (see SM section~\ref{SM-sec:FT}). 
Still, further sources of inaccuracy
could be ascribed to Co-HM intermixing at the interface,
or to the particular moir\'e pattern between the Gr and Co unit cells
or to the shape anisotropy (SA), which has been neglected so far.

Indeed, some degree of compositional disorder may well be present close
to the Co/HM interface~\cite{bib:ajejas20}, as can be inferred from the
variable intensity at some interface atomic planes in the HR-STEM 
images in Figs.~\ref{fig:sample_characterization} and 
\ref{SM-fig:microstr}.
As shown in the SM section~\ref{SM-sec:fccvshcp}, Pt-Co mixing at the early stages
of the Co growth is a stabilizing factor towards the $fct$ structure of the 
films versus the energetically favoured $hcp$ stacking. Still,
interfacial mixing should have a minor effect in the MAEs of thick films
once the bulk regime has been reached.

The influence of the precise Gr/Co interface geometry on the MAE has been 
addressed by performing analogous calculations for 
(1$\times$1)-Co\n/HM$_{12}$ slabs, i.e., 
without the Gr capping layer on top, as well as for
Gr/Co\n/Pt$_5$ slabs assuming two standard Pt/Gr moir\'e patterns.
Results for these scenarios are presented in the SM section~\ref{SM-sec:moire}. It turns
out that large deviations in the MAEs of up to 0.5~meV or even larger are always
constrained to the ultra-thin limit (n$\leq 4$), in analogy with the case of
unstrained Co multilayers with $hcp$ stacking~\cite{bib:yang15}.
As n increases, the relative contribution of the vacuum/Co or Gr/Co interfaces 
is reduced and the MAE of the films also approaches the bulk Co $fct$ limit.
We have accounted for the uncertainty on the MAEs due to the precise
moir\'e pattern in Fig.~\ref{fig:mae} via large error bars
that decrease with the Co thickness as interface contributions become
less relevant. However, even after considering such ample errors, the 
theoretical critical thicknesses hardly change and remain considerably smaller
than the experimental ones.

Concerning the shape anisotropy, as shown in the SM section~\ref{SM-sec:sa}, the 
SA of ideal (monodomain) $fct$ 
Gr/Co\n/HM films favors in-plane magnetization,
i.e., negative MAE values, with an energy contribution that evolves almost 
linearly with n and, therefore,
its incidence on the total MAEs would be to even further reduce the PMA value.
Nevertheless, we have excluded the SA term in Fig.~\ref{fig:mae}
since, in real samples, the existence of multiple domains
with different/opposite magnetization directions during magnetization reversal~\cite{bib:olleros19} should
reduce considerably its contribution.

{\bf Stacking defects in the $fct$ films.} The above analysis demonstrates that perfect $fct$ Co
films cannot hold PMAs at large Co thicknesses. However, from the STEM images
shown in Figs.~\ref{fig:sample_characterization} and \ref{SM-fig:microstr} structural defects in 
the grown samples are common, mainly in the form of SFs or TBs.
The appearance of a TB (SF) requires the $hcp$ stacking of one 
(two) Co layers, --ABC--A/B/A--ABC-- (--ABC--A/B/A/B--CAB--), and therefore, 
incorporates an additional interface
into the film which could well delay reaching the bulk $fct$ limit --indeed, 
this type of defect is known to significantly alter the band structure 
of the Co film~\cite{bib:vazquez00}.
We have addressed this possibility by inserting a single TB in the 
Gr/Co\n/HM$_{12}$ films. In Figs.~\ref{fig:mae}(C) and (D) we have included
the resulting MAEs when the TB (that is,
the $hcp$ stacked layer) is incorporated at the third 
(red dots and lines) or fourth (blue) Co atomic plane starting from the Co/HM 
interface (see panel B in the same figure). Remarkably, 
in both cases the appearance of the
TB yields a large shift of around 1~meV in the MAE towards PMA when
compared to the defect-free case (black dots). After 
this initial large jump, the film again approaches the bulk
$fct$ limit but this time with an additional and rather 
large interfacial contribution associated to the TB, which
shifts the critical thicknesses of the films up to the 
10~ML range, in better agreeement with the experiments. 

Table~\ref{tab:tb10} summarizes the interfacial contributions to the MAE for a 
TB appearing in any of the Co layers in a (1$\times$1)-Gr/Co$_{10}$/HM$_{12}$ 
slab. They always attain positive 
values, well above 0.6~meV in many cases, although no clear trend with the TB
location can be envisaged due to interference effects between the three
interfaces. In the case of Ir, 
where the experimental switch of the MAE occurs at around \nc=10~MLs, a
single TB (on average) across the $fct$ Co film would be sufficient to 
overcome the in-plane bulk-like contribution of the ideal $fct$ film
(see dark line Fig.~\ref{fig:mae}(D)). In the case of Pt, where the critical 
thickness is as large as \nc=20~MLs, the required out-of-plane TB contribution 
would be of the order of
1.3~meV, corresponding to around two TBs on average throughout the film. 
Therefore, stacking defects in the Co $fct$ films represent a robust and 
necessary ingredient to achieve PMAs at large thicknesses. 
Such a large interfacial contribution may be understood from the fact that a 
strained Co $hcp$ crystal presents large PMAs of 0.49 and 0.56~meV/atom at the
Pt and Ir in-plane lattice constants, respectively (dark blue lines in 
Figs.~\ref{fig:mae}(C) and (D)), and the presence of a TB can be considered as an initial 
stage towards an $hcp$ stacking. Furthermore, since SFs comprise two 
consecutive locally stacked $hcp$ layers, their PMA contribution is expected
to be larger than that of a single TB. As a representative example, a SF placed
at the second and third Co layers in a Gr/Co$_4$/Pt$_{12}$ film yields a PMA 
more than twice larger than if only a single TB is considered (1.27~meV versus 
0.55~meV, respectively).

\begin{table}
\caption{Difference in the calculated MAE (in meV) between a 
$(1 \times 1)$-Gr/Co\n/HM$_{12}$
slab with a TB located at layer $i = 2-9$ from the 
Co/HM interface (Fig.~\ref{fig:mae}(B)) and that of the 
same slab without defects (ideal Co-$fct$).
\label{tab:tb10}}
\begin{tabular}{cccccccccc}
\hline \hline
 i & 1 & 2 & 3 & 4 & 5 & 6 & 7 & 8 & 9  \\
\hline
Pt & 0.11 & 0.37 & 0.55 & 0.62 & 0.68 & 0.66 & 0.69 & 0.47 & 0.75 \\
Ir & 0.81 & 0.57 & 0.77 & 0.42 & 0.49 & 0.53 & 0.62 & 0.35 & 0.74 \\
\end{tabular}
\end{table}

\begin{figure}
\includegraphics[width=0.9\columnwidth]{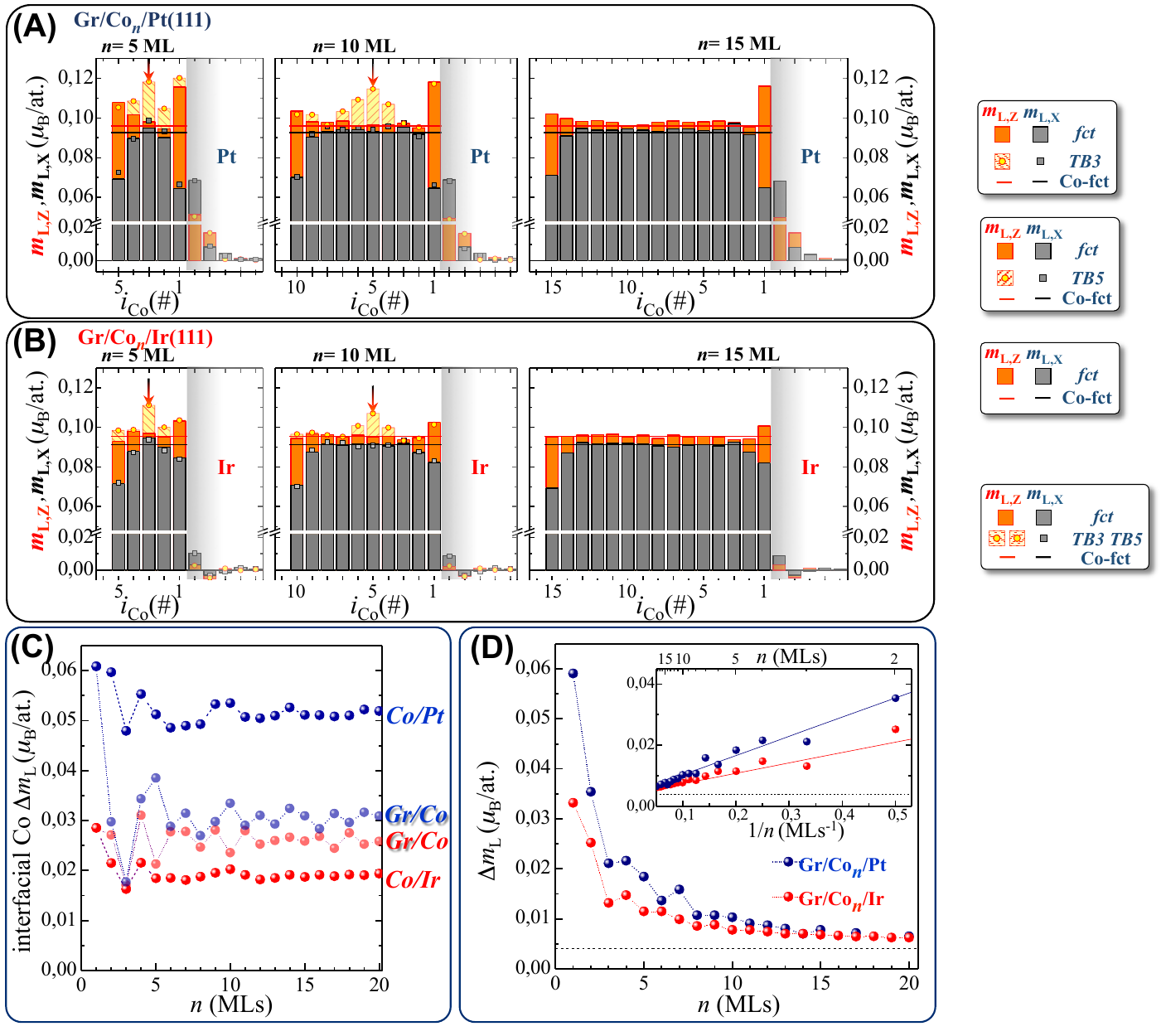}
\caption{\label{fig:omm}
{\bf Calculated layer-resolved orbital
moments in Gr/Co/HM heterostructures} for 
(A) ($1\times1$)-Gr/Co$_\mathrm{n}$/Pt$_{12}$ and 
(B) ($1\times1$)-Gr/Co$_\mathrm{n}$/Ir$_{12}$ slabs and for indicated $\mathrm{n}$. 
Red (black) bars correspond to out-of-plane $m_{L,Z}$ (in plane $m_{L,X}$) 
projections of a defect-free slab with a Co $fct$ sequence. 
Bars with circle (square) symbols indicate the
corresponding values when a TB defect is introduced at the third ($i_\mathrm{Co}=3$)
and fifth ($i_\mathrm{Co}=5$) Co planes for the n=5 and 10 slabs,
respectively. Solid lines refer to the bulk values. 
(C) Orbital magnetic moment anisotropy ($\Delta m_L=m_{L,Z}-m_{L,X}$) 
of the interfacial Co atoms as a function of the Co
slab thickness n, for Pt (blue) and Ir (red) buffer 
layers. The light and dark lines refer to the Gr/Co and Co/HM
interfaces, respectively. 
(D) Effective orbital magnetic
moment anisotropy ($\Delta m_L^{eff}$) as a function of the Co thickness for the
indicated heterostructures. The dotted lines indicates the
averaged value of the inner Co layers. The inset shows that
$\Delta m_L^{eff}$ follows an inverse law with n, i.e., 
interfacial nature, whereas the slope is twice in the case of
Gr/Co/Pt.} 
\end{figure}

{\bf Orbital Magnetic Moments.} 
We next turn our attention to the 
orbital magnetic momments (OMMs) in the Co films.
Figures~\ref{fig:omm}(A) and (B) show the OMMs projected on individual
Co and first interfacial HM atoms for a Pt$_{12}$ and Ir$_{12}$ buffer 
layer, respectively, calculated along the out-of-plane, $m_{L,Z}$, and in-plane, 
$m_{L,X}$, spin quantization axes. We include the layer resolved 
OMMs for two thin n$=3,5$ and a thick, n$=10$, Co films
(equivalent data for other thicknesses are presented in SM Fig.~\ref{SM-fig:omms}).
If we first focus on the ideal $fct$ films 
and the thick n$=10$ limit, both $m_{L,X}$ and $m_{L,Z}$ remain fairly constant across 
the Co layer, with clear 
deviations only at the upper and lower interfaces.
The $m_L$ values at the inner layers of the slabs 
approach those of their respective bulk 
$fct$ phases (SM Table~\ref{SM-tab:omma_contrib_fleur}), 
confirming that the bulk limit has been reached at these thicknesses.
At the interfaces, the Co OMMs behave differently depending on the spin
quantization axis; while $m_{L,X}$ is strongly reduced at both sides of the
Co layer, $m_{L,Z}$ shows an increase in most of the cases.
Hence, the calculated orbital magnetic moment anisotropy (OMMA),
defined as $\Delta m_L=m_{L,Z}-m_{L,X}$, remains positive across the entire Co film with
a marginal value of $\sim$0.004~$\mu_B$ at the inner layers, but presenting
an enhancement of one order of magnitude at the interfaces. 
When one considers the thinner slabs, n$=3,5$, 
similar conclusions can be drawn for both interfaces 
and both HMs, except that now the film 
is not thick enough to develop a plateau at the central layers.

In panel C we present the evolution of the OMMAs projected on the first
(Co$_{\mathrm{Gr}}$) and last (Co$_{\mathrm{Pt,Ir}}$) Co layers as a function
of the film thickness. 
At the Gr/Co side, and apart from strong oscillations in the
n$\leq 5$ range specially for Pt (light blue line), 
$\Delta m_L$ remains fairly constant around 0.03~$\mu_B$ and slightly
smaller for Ir (light red line). It is interesting
to note that, despite the OMMs projected on the C atoms are negligible, 
such large OMMA is induced by the Gr layer. As shown in the SM Fig.~\ref{SM-fig:omms},
when the Gr capping layer is removed the projections of $m_{L,X}$ and $m_{L,Z}$ at the 
first Co layer both show a similar enhancement of up to $\sim 0.13$~$\mu_B$, 
but their difference $\Delta m_L$ becomes negligible. Hence,
the Gr layer introduces a highly localized OMM anisotropy at the top of the Co film, 
otherwise absent, by strongly reducing $m_{L,X}$ compared to a smaller decrease of 
$m_{L,Z}$, the effect being independent of the nature of the HM buffer layer.
On the other hand, at the lower Co/HM interface there exist clear differences 
between the two HMs (dark blue and red lines in Fig.~\ref{fig:omm}(C).
For Co$_{\mathrm{Ir}}$ we find an interfacial OMMA of $\sim 0.02$~$\mu_B$, 
while the OMMs in the Ir buffer drop to very small values, even marginally
negative, already at
the first layer. In the Pt case the OMMA at the Co$_{\mathrm{Pt}}$ layer is 
considerably larger ($\sim 0.05$~$\mu_B$), while
this time the polarization of the ferromagnetic Co penetrates into 
the Pt buffer as the OMMs of the first Pt layer are around 0.05-0.07~$\mu_B$
and, notably, present a negative OMMA. We attribute this unexpected
proximity effect to a larger induced SOI compared to the Ir 
(see SM section~\ref{SM-sec:FT}) as well as to the fact that the Pt layer shows a larger
induced magnetic moment at the interface compared to Ir 
(0.25 versus 0.11~$\mu_B$, the former value being in good agreement 
with experimental values observed in Co/Pt interfaces \cite{bib:ferrer89}). 
In the same line, we recall that
Belabbes~{\it et al.}~\cite{bib:belabbes16} also found clear differences 
in the Dzyaloshinskii-Moriya interaction strength between the Co/Pt(111) and 
Co/Ir(111) systems.

Last, we address the OMMs associated to a TB. The red circles in
Figs.~\ref{fig:omm}(A) and (B) correspond to the OMMs for defected Co films;
in the n$=5$ case the TB has been incorporated at the third ($i_\mathrm{Co}=3$) Co layer, while
in the thicker n$=10$ slab at the fifth ($i_\mathrm{Co}=5$) layer. 
Notably, in all cases, $m_{L,Z}$ shows a pronounced peak precisely at the location
of the $hcp$ stacked layer, whereas $m_{L,X}$ remains essentially unchanged. Since
the same behavior is reproduced for all other locations of the TB in the Co
film (not shown), we conclude that an $hcp$ stacked layer consistently yields 
a localized increase of the OMMA.

In order to compare the theoretical OMMAs against the X-ray absorption data
shown in Fig.~\ref{fig:mokexmcd}(B), 
we define an effective OMMA for the Co atoms in each slab as
$\Delta m_L^{eff} = \frac{1}{{\mathrm n}} \sum_{i=1}^{{\mathrm n}} (m^i_{L,Z} - m^i_{L,X})$, 
where the index $i$ runs over the Co atomic planes.
Fig.~\ref{fig:omm}(D) shows the resulting values in the defect-free case as a 
function of thickness for both HM buffer layers.
The two curves show a rapid decrease with the Co thickness governed by the 1/n
factor. Indeed, the $\Delta m_L^{eff}$ data points can be linearly fitted 
(see inset), with slopes following the experimental trend
shown in the inset of Fig.~\ref{fig:mokexmcd},
namely around twice larger in the case of the Pt buffer.
For each thickness, the main cointributions to the effective OMMA are those 
of the Co$_{\mathrm{Gr}}$ and Co$_{\mathrm{HM}}$ interface atoms
(highlighted in Fig.~\ref{fig:omm}(D)) as well as those displaying a local
$hcp$ stacking in the defected slabs.
Overall, the theoretical effective OMMA
curves are in reasonable qualitative agreement with the XMCD data (see 
Fig.~\ref{fig:omm}(D) and the inset of Fig.~\ref{fig:mokexmcd}).
Still, there are relevant differences which deserve a further discussion. 

First, the calculated OMM values are significantly underestimated
by a factor of around three with respect to the experimental ones. 
Even after the inclusion of TBs close to the surface,
which lead to an overall increase of the $\Delta m_L^{eff}$ values (not
shown), a significant experiment-theory difference remains. 
We ascribe it to the inherent limitation of DFT in the treatment of
the electronic correlations~\cite{bib:keshavarz15}, which is well
known to underestimate (overquench) OMMs of isolated magnetic atoms on 
surfaces~\cite{bib:baumann15,bib:gallardo19}, 
and has also been noted, for instance, in 
ultra-thin magnetic films~\cite{bib:Lehnert10}.

Second, the XCMD results point to a change of sign
in $\Delta m_L$ at similar critical thicknesses at which the MAE also switches.
At contrast, the calculated OMMAs remain always positive
at every Co plane and for all thicknesess considered, regardless if the MAE is
out-of- or in-plane.
\footnote{In an approximation that takes into account only the
atomic orbital spatial distribution, as it is the case of the present calculations
(atomic OMM values are obtained from projections of the Kohn-Sham wavefunctions on atomic orbitals), 
the $d_{x^2-y^2,xy}$ orbitals contribute to $m_{L,Z}$, while $d_{z^2}$
and $d_{xz,yz}$ to $m_{L,X}$~\cite{bib:stoehr99}.  
Since the latter in-plane contributions (corresponding to negative OMMA values)
are attenuated in a bulk-like environment due to hybridization between atomic planes,
the net result is a positive OMMA (see SM section~\ref{SM-sec:fleur}).}
We recall that a direct relationship between the MAE and the OMMAs, as the 
experiments here suggest, is far from having a trivial 
explanation~\cite{bib:bruno89,bib:vanderlaan97,bib:gambardella03,bib:blanco19,bib:Andersson07,bib:Lehnert10}.  
As a matter of fact, despite the strong SOI conferred by Pt and Ir to the band structure,
the correlation between the two properties is not guaranteed, as shown, for instance, in
$3d-5d$ magnetic alloys~\cite{bib:blanco19} and thin-film
heterostructures~\cite{bib:Andersson07,bib:Lehnert10}.

From a theoretical point of view, the OMMA and the MAE are expected to be 
proportional only under the following conditions~\cite{bib:vanderlaan98}:
the SOI strength must be small enough for the MAE to be treated as 
a second-order perturbation effect on the electronic wavefunctions, 
the anisotropic part of the spin distribution (quadrupole contribution) 
must be negligible and the orbital ground state must be non degenerate. 
The strained Co $fct$ bulk limit already represents an example where
any of the above conditions is not fullfilled, as the DFT-derived in-plane 
MAE is associated with an out-of-plane OMMA (see SM section~\ref{SM-sec:fleur} 
for a more detailed discussion).
Finally, it is also timely to recall that the formulation of the sum rules, 
based on an atomic orbital picture, typically employed to obtain orbital 
magnetizations from the XMCD data~\cite{bib:vdL92,bib:thole92,bib:stoehr99} yield, 
instead of an orbital magnetization that accounts for the effect of the 
itinerant electron many body wave function, an effective orbital moment that 
has a great practical importance for interpreting XMCD spectra~\cite{bib:kunes00,bib:souza08,bib:resta20}.

\section{Summary and Conclusions}
\label{sec:conclusion}

The magnetic anisotropy of Gr/Co\n/HM(111) (Gr=graphene, HM=Pt,Ir) heterostructures,
grown by thermally activated Co intercalation,
has been characterized experimentally by MOKE and XMCD
with focus on the evolution of the magnetic properties with Co thickness.
This growth technique produces a pseudomorphic film with ABC stacking ($fct$ structure), 
which shows perpendicular magnetic anisotropy (PMA) up to 20 and 10~MLs on 
Pt and Ir substrates, respectively, as revealed by MOKE measurements of 
the perpendicular remanent magnetization $M_{Z,R}$. 

In order to rationalize these findings, extensive first-principles 
DFT calculations including the SOI self-consistently have been performed for
(1$\times$1)-Gr/Co\n/HM$_{12}$ slabs with up to n$=20$ Co layers. 
Such unusually thick slabs were found necessary to properly
characterize the transition from thin Co films to the bulk $fct$ limit at around 
a Co thickness threshold value of 8~ML, where interface effects and finite size quantum oscillations start to become negligible.
Beyond this threshold, the Co film approaches the 
bulk Co $fct$ limit and shows an increasing in-plane anistropy as the 
film grows thicker. Hence, we find that in the ideal $fct$ system the PMA, 
i.e., positive MAE values, would vanish at much lower Co thicknesses than those experimentally observed. 
However, we have proved that the inclusion of structural defects in the form of
twin boundaries to mimic a local $hcp$ stacking in the Co film, actually
seen in real samples mainly as stacking faults, can significantly
delay the MAE switching and explain the observed critical thicknesses.
Therefore, we reach the counterintuitive conclusion that it is precisely the existence of structural defects what prevents PMA degradation in these, otherwise almost
perfect, heterostructures.

A sum rule analysis of the XMCD spectra shows a sizable 
orbital magnetic moment anisotropy $\Delta m_L$ of the Co atoms that
switches from out-of-plane to in-plane at similar Co thicknesses 
as $M_{Z,R}$ is observed to vanish in MOKE. 
DFT predicts a perpendicular OMMA, regardless of the presence or absence of 
stacking defects in the Co film, which shows an attenuation 
with increasing Co thickness compatible with the observations.
Importantly, the calculations also reveal that this behaviour is 
dominated by the large OMMA at the Gr/Co interface, and which
is absent at the vacuum/Co interface.

\section{Methods}
\label{sec:methods}


\subsection{Experiments}

{\bf Sample Preparation.} 
The epitaxial Gr-based epitaxial heterostructures were grown in ultrahigh-vacuum (UHV) 
condition on commercially available SrTiO$_3$(111)- and Al$_2$O$_3$(0001)-oriented 
oxide single crystals. The oxide crystals were ex/situ annealed in air at 1370\,K for 2\,h 
in order to obtain flat surfaces with large terraces prior to their insertion in the UHV chamber. 
Epitaxial (111)-oriented Pt and Ir buffers with thicknesses ranging from 10 to 30\,nm were 
deposited by DC sputtering in $8 \times 10^{-3}$\,mbar Ar partial pressure at 670\,K with a 
deposition rate of 0.3\,{\AA}\/s. The quality of the fabricated Pt and Ir templates resembles 
the one of a single crystal, as demonstrated by LEED and XPS surface analyses.

The Gr monolayer was generated on Pt (Ir)/MgO(111) templates by exposing the samples 
kept at 1025\,K in UHV ($1 \times 10^{-9}$\,mbar) to ethylene gas at a partial pressure of 
$2 \times 10^{-8}$\,mbar for 30\,min. The Gr/Pt/oxide(111) sample was cooled down to RT and Co 
was deposited on the top by e-beam evaporation at RT with a deposition rate of 0.04\,{\AA}\/s. 
The sample was gradually heated up to 550\,K while acquiring XPS spectra to verify in real 
time the intercalation of Co underneath the Gr sheet. Once the intercalation was completed, 
the resulting sample was Gr/Co\n/Pt/oxide(111). 

{\bf High-Resolution STEM.} 
Electron microscopy observations were carried out in a JEOL ARM200cF microscope equipped with 
a CEOS spherical aberration corrector and a Gatan Quantum EEL spectrometer at the 
Centro Nacional de Microscop\'{\i}a Electronica (CNME) at the University Complutense of Madrid. 
Specimens were prepared by conventional methods, including mechanical polishing and Ar ion milling. 

{\bf Polar Kerr Magnetometry and Microscopy.} 
The RT vectorial-Kerr experiments were performed in polar configuration by using p-polarized light 
(with 632\,nm wavelength) focused on the sample surface and analyzing the two orthogonal 
components of the reflected light. This provides the simultaneous determination of the hysteresis 
loops of the out-of-plane and in-plane magnetization components, that is, $M_Z$ and $M_X$, 
by sweeping the magnetic field along the sample out-of-plane ($\hat z$) direction.

{\bf XAS-XMCD.} 
The XAS and magnetic circular dichroism experiments were carried out at the BOREAS beamline of the 
ALBA synchrotron using the fully circularly polarized X-ray beam produced by an apple-II type 
undulator~\cite{bib:barla16}.  
The base pressure during measurements was $\sim 1 \times 10^{-10}$\,mbar. 
The X-ray beam was focused to about $500 \times 500$\, $\mu$m$^2$, and a gold mesh has been used 
for incident flux signal normalization. The XAS signal was measured with a Keythley 428 current 
amplifier via the sample-to-ground drain current (total electron yield TEY signal). The magnetic 
field was generated collinearly with the incoming X-ray direction by a superconducting vector 
cryomagnet (Scientific Magnetics). To obtain the spin averaged XAS and the XMCD, the absorption spectra were 
measured as a function of the photon energy both for parallel and antiparallel orientation 
($\mu^+(E)$ and $\mu^-(E)$) of the photon spin and the magnetization of the sample. 
We recall that such XMCD measurements at the Co and $L_{2,3}$ absorption edges provide direct 
element-specific information on the magnitude and sign of the projection of Co magnetizations along 
the beam (and field) direction. 

\subsection{Calculations}

DFT calculations have been carried out with the GREEN code~\cite{bib:cerda97,bib:rossen13}
and its interface to the \textsc{Siesta} DFT-pseudopotential 
package~\cite{bib:soler02} using the PBE exchange and correlation functional~\cite{bib:pbe96} 
and the fully-relativistic pseudo-potential (FR-PP) 
approach~\cite{bib:cuadrado12} to include the SOI self-consistently.
The basis set consisted of strictly localized atomic orbitals generated following
a double-zeta scheme for all atoms and employing a confinement energy 
(Energy Cut-off) of 100~meV. Pseudo-core corrections were included for the metal
atoms in order to describe accurately magnetic and SOI-derived 
properties~\cite{bib:bloechl94}. An electronic temperature 
$kT=20$~meV was used for the Fermi-Dirac distribution
function and ultra fine $k$-space grids of at least 75$\times$75 relative to the
(1$\times$1)-HM lattice together with real space meshes with a 
resolution $\sim 0.04$~\AA$^3$ (equivalent to Mesh Cut-offs between 1,000 and 2,000 Rydbergs)
were employed to ensure a poper convergence, within
less than 0.02~meV, in all reported MAEs. 

Although the results presented in the main text correspond to 
(1$\times$1)-Gr/Co\n/HM$_{12}$ model structures consisting of a 12 layers thick  
$fcc$ HM buffer layer (111) oriented with a varying number of Co layers following 
the $fcc$ stacking sequence on top, n$=1-20$, plus a capping ($1\times1$)-Gr 
layer (see Figure~\ref{fig:mae}(A)), we additionally considered
alternative models including structural defects such as TBs, 
varying the buffer layer thickness or considering
different moir\'e patterns at the Gr/Co interface (a detailed description of
all of them together with their most relevant structural parameters after
the atomic relaxations, as well as tetragonally distorted Co $fct$ bulk phases
are provided in the SM section~\ref{SM-sec:allmodels}).

The magnetic anisotropy energy (MAE) is defined here as:
\begin{equation}
\mathrm{MAE} = E_{tot}^x - E_{tot}^z 
\end{equation}
where $E_{tot}^{x,z}$ stand for the total energies, including 
SOI terms fully self-consistently, for spins aligned along the $OX$ and $OZ$ 
axes, so that PMA corresponds to a positive value of the MAE.

The MAE is a property in the meV and sub-meV range, extremely 
sensitive to calculation parameters, particularly the basis size. 
In order to obtain accurate MAE values, the convergence of this 
quantity with calculation parameters (energy cut-off, 
reciprocal space sampling, smearing of the Fermi level) has been carefully checked.
Furthermore, we have examined if the force theorem approach~\cite{bib:weinert85,bib:li90,bib:daalderop90}
to obtain MAE values is a reliable method for the present systems. 
These tests are gathered in the SM section~\ref{SM-sec:FT}.

In order to cross-check the accuracy of the SOI-derived properties calculated 
under the FR-PP approximation used in the \textsc{Siesta-Green} code, we have performed 
selected benchmark calculations using the DFT full-potential linearized augmented 
planewaves (FLAPW) formalism~\cite{bib:krakauer79,bib:wimmer81}, 
as implemented in the FLEUR code~\cite{bib:fleur}. 
Same as for \textsc{Siesta-Green}, we used the PBE exchange and correlation  
functional~\cite{bib:pbe96} while the SOI was  included  fully  self-consistently~\cite{bib:li90}. 
The  FLAPW  basis  set is  constructed  with  sufficiently  fine Monkhorst-Pack-point 
meshes~\cite{bib:monk76} to sample the first Brillouin zone and used plane wave expansion 
cut-offs of 4\,a.u. for the wavefunctions, and 12\,a.u. for the density and potential.
For the local basis, the Co $4s,3p,3d$ electrons were treated as valence states and the $3s$  
as a local orbital. The partial wave expansions were constructed with a $l_{max}=8$  
cut-off in a muffin-tin  sphere  of  radius  of 1.2\,{\AA}.  The  Fermi  energy was 
determined  by  smearing  with a Fermi-Dirac function of $kT=14$\,meV. 
In particular, the FLEUR code was employed for the calculation of the MAEs 
and orbital magnetic momenta values of a strained free-standing Co monolayer 
and the bulk $fct$ and strained $hcp$ Co limits. The comparison versus the 
\textsc{Siesta-Green} values, shown in the SM 
Tables~\ref{SM-tab:mae_omma_fleur}, \ref{SM-tab:omma_contrib_fleur}, 
provides an excellent agreement for the OMMs in all cases 
in spite of the use of a different basis, as well as for the MAEs in the bulk phases. 
A discrepancy of 0.7~meV is however found for the MAE of the free-standing monolayer.

\begin{acknowledgements}

Discussions with J.J. Saenz (Mole) and Raffaele Resta are kindly acknowledged.
Financial support from MINECO [grant numbers RTI2018-097895-B-C41, RTI2018-097895-B-C42
and RTI2018-097895-B-C43 (FUN-SOC),
PID2019-103910GB-I00, FIS2016-78591-C3-1-R and FIS2016-78591-C3-2-R(SKYTRON),
PGC2018-098613-B-C21 (SpOrQuMat), PCI2019-111908-2 and PCI2019-111867-2 (FLAGERA 3 grant SOgraphMEM)],
from Regional Government of Madrid (grant number P2018/NMT-4321 (NANOMAGCOST-CM)) and
from Gobierno Vasco-UPV/EHU (grant numbers GIU18/138 and IT-1246-19).
We acknowledge experiments at ALBA BL29 via proposal \#2019023333.
IMDEA-Nanociencia acknowledges support from the ``Severo Ochoa'' Program for
Centres of Excellence in R\&D (MINECO, Grant SEV-2016-0686).
Computational resources were partially provided by the DIPC computing center.

\end{acknowledgements}

\bibliography{grcoptir}

\end{document}